# Distribution of $^7$Li Implants in Nb Films for a Sterile Neutrino Search

Hendrik Hadenfeldt[a], Jonas Arlt[a]*[1], Tobias Meyer[a], Felix Junge[b], Stephan Friedrich[c], and Cynthia A. Volkert[a]

[a]*Institute of Materials Physics, University of Göttingen, Göttingen, Germany;*
[b]*2nd Institute of Physics, University of Göttingen, Göttingen, Germany;*
[c]*Lawrence Livermore National Laboratory, Livermore, California, USA*

* Corresponding author: jonas.arlt@tu-berlin.de

[1] Now at: Institute of Materials Science and Technologies, TU Berlin, 10587 Berlin, Germany

# Abstract

The BeEST experiment is measuring the $^7$Li recoil spectrum from the decay of $^7$Be implanted into Ta-based sensors to provide the most stringent limits on the existence of sterile neutrinos in the sub-MeV mass range. Its sensitivity is limited by spectral broadening due to interactions between the atomic shell of the $^7$Be/$^7$Li and the Ta sensor film. This study is investigating the suitability of Nb as an alternative sensor material by scanning transmission electron microscopy (STEM) and atom probe tomography (APT) of $^7$Li-implanted Nb films. STEM does not detect any change in the microstructure of polycrystalline Nb due to $^7$Li implantation. APT reveals some segregation of $^7$Li at grain boundaries in Nb. While this may slightly alter the $^7$Li binding energies, the effect on the recoil spectrum is not expected to be strong enough to preclude using Nb-based detectors in future phases of the BeEST experiment.

# Keywords:

# Impact statement:

Atom probe tomography revealed lithium segregation to niobium grain boundaries, while STEM showed no detectable change in Nb microstructure, suggesting minimal impact on BeEST experiment sensitivity.

# Introduction

The Standard Model of particle physics (SM) is a highly successful theory of subatomic particles and their interactions. Its predictions have been tested and confirmed many times under increasingly stringent conditions. But despite its remarkable success, it is known to be incomplete, as it cannot account e.g. for dark matter or the matter-antimatter asymmetry in the Universe [1]. The observation of neutrino oscillations and thus finite neutrino masses is currently the only confirmed violation of the SM in its original form [2,3]. In addition, neutrinos are the only particles of the SM with an intrinsic chirality, in that they have only been observed as left-handed. Most extensions of the Standard Model therefore postulate the existence of right-handed so-called sterile neutrinos that are inactive under weak interactions and interact with SM particles only through gravity and mixing [4]. These sterile neutrinos can theoretically exist at any mass scale and are currently the subject of several experiments worldwide [5].

The *Be*ryllium-7 *E*lectron capture in *S*uperconducting *T*unnel junctions (BeEST) experiment currently provides the most stringent limits on the existence of sterile neutrinos in the sub-MeV mass range [6]. It uses high-resolution superconducting tunnel junction (STJ) sensors to accurately measure the recoil energy of $^7$Li from the electron capture decay of $^7$Be that has been implanted into the STJs. Since electron capture is a two-body process and the neutrino escapes from the STJ detector, the $^7$Li recoil energy is (in principle) monochromatic and has a value of 56.826(9) eV for active neutrinos with masses below 1 eV. The emission of a sterile neutrino in the sub-MeV mass range would reduce this recoil energy and produce additional peaks in the spectrum as a signature.

The electron capture decay of $^7$Be produces a hole in the electronic shell of the $^7$Li daughter that decays on a ~ns time scale and adds its energy to the recoil energy. In addition, roughly 10% of all $^7$Be decays produce the excited nuclear state $^7$Li* that decays with a half-life of 73 fs. The $^7$Li recoil spectrum therefore consists (in practice) of four separate peaks, two for K capture and two for L capture and decay into the ground state and excited state of $^7$Li, respectively (**Fig. 1**) [7]. Surprisingly, the primary K-capture peak into the $^7$Li ground state at ~106 eV has a width of ~6.7 eV FWHM that significantly exceeds the energy resolution of the STJ detector of 2 eV FWHM. This degrades the minimum detectable peak shift and therefore limits the sensitivity of the BeEST experiment for sterile neutrino masses below ~100 keV. At least some of that width is due to variations of the Li 1s binding energy at different lattice sites in the polycrystalline

Ta absorber film of the STJ detector [8]. The observation also raises the question which other material-dependent effects might be present in Ta-based STJs that affect the detector response and introduce systematic uncertainties.

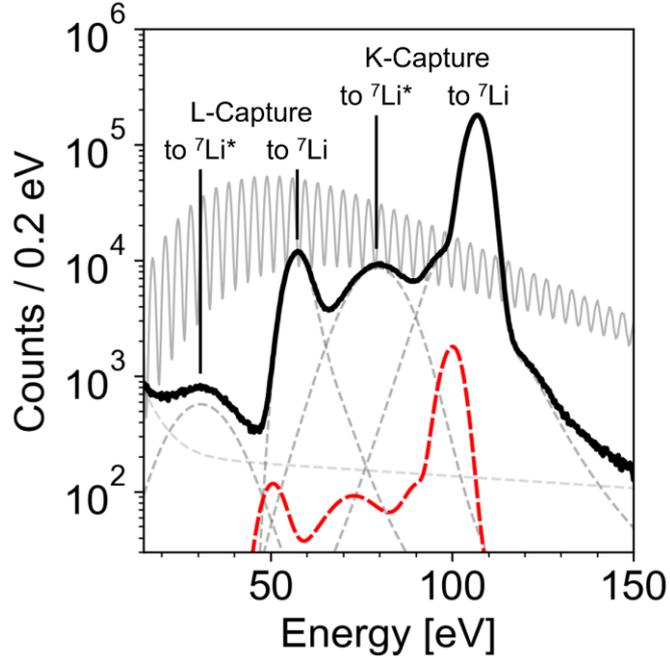

**Figure 1. $^7$Li recoil spectrum (black)** showing four peaks for the different $^7$Be to $^7$Li decay channels and a broad background due to gamma radiation interactions in the STJ detector substrate (dashed). The spectrum generated by a hypothetical 300 keV sterile neutrino signal is shown in red (dashed), along with the laser calibration signal (grey). Reproduced from data given in [6].

Niobium is an attractive alternative material for STJ fabrication since Nb-based STJ sensors can also be fabricated with an energy resolution around ~2 eV FWHM at energies of interest for the BeEST experiment [9]. Nb-STJs would also allow separating material-dependent background effects from a possible sterile neutrino signal, provided that $^7$Be implantation and decay to $^7$Li do not damage the Nb absorber film or unacceptably broaden the detector response.

In this study, we investigated the impact of Li implantation on Nb films relevant for future Nb-based STJs. Using 4D scanning transmission electron microscopy (4D-STEM), we assessed the microstructure of the Nb films before and after implantation, while atom probe tomography (APT) provided insights into Li distribution within polycrystalline Nb.

## Materials and Methods

*Materials*

We investigated Nb films sputter-deposited by STAR Cryoelectronics LLC under the same conditions as those used for absorbers in Nb-based STJs, which are potential candidates for future phases of the BeEST experiment. The samples consisted of a single ~280 nm thick Nb layer deposited on an oxidized silicon wafer.

To enable a direct comparison between lithium-implanted and non-implanted Nb regions within the same sample, we used nanolithography to create an implantation setup. A grid with sharp edges was patterned by electron-beam lithography at the Institute for X-ray Physics, Göttingen. Gold was chosen as a blocking layer due to its high stopping power, with SRIM simulations [10] confirming that a 210 nm thick layer effectively halts $^7$Li ions at 30 keV (see supplementary **Fig. S1a**). To ensure complete blocking, we deposited a 270 nm thick gold layer. This approach allowed us to study the structural effects of $^7$Li implantation on the Nb matrix while eliminating systematic differences from sample preparation and analysis.

Lithium implantation was performed at the 2nd Institute of Physics, Göttingen, using the ADONIS particle accelerator [11]. The ions were implanted at 30 keV with a fluence of $5\times10^{15}$ atoms/cm$^2$, determined by integrating the electric current on the sample during implantation. This corresponds to an average lithium content of 0.5 at.% over a maximum implantation depth of 200 nm in Nb (see supplementary **Fig. S1b**). The implantation process lasted 57 min with an ion current ranging from 0.4 to 0.8 µA.

*Conventional and scanning transmission electron microscopy*

The TEM specimen was prepared using an FEI Helios G4 DualBeam Ga-focused ion beam (FIB) in conventional cross-section geometry. Thinning windows were milled approximately 3 µm from the gold edge on both the gold-covered and uncovered regions. This setup enables the analysis of both non-implanted regions and regions containing the full $^7$Li implantation dose within a single TEM lamella (see **Fig. 2**). The final FIB milling was performed at an acceleration voltage of 2 kV.

The implanted and non-implanted regions were analyzed using an FEI Titan 80-300 E-TEM G2 operated at 300 kV. Four-dimensional STEM (4D-STEM), including virtual dark field imaging, and electron energy loss spectroscopy (EELS) were employed. 4D-STEM data was acquired with a probe current of 30 pA and a semi-convergence angle of 1 mrad, using a Gatan Ultrascan 1000XP camera. A custom script synchronized beam movement and image acquisition [12]. A camera binning of 8 was used, resulting in a sampling of 0.136 nm$^{-1}$ per pixel. EELS data was collected with a probe current of 100 pA and a semi-convergence angle of 10 mrad using a Gatan GIF Quantum 965 ER spectrometer equipped with an Ultrascan 1000XP camera. The dispersion was set to 0.1 eV per channel.

*Atom probe tomography*

APT specimens were prepared using an FEI Helios G4 DualBeam Ga FIB following a lift-out procedure based on Ref. [13]. In brief, a lamella was extracted from the bulk sample, horizontally attached to a pre-prepared tungsten tip, and shaped into the final APT tip by lateral milling. This ensured that the grain boundaries of the columnar Nb film were primarily perpendicular to the APT measurement direction (see **Fig. 2b**), where APT provides the highest spatial resolution [14]. Before APT analysis, the final tips were characterized using TEM (FEI Titan ETEM G2) operated at 300 kV.

APT experiments were performed at the Institute for Materials Science, Stuttgart, using a custom-built wide-angle laser-assisted atom probe system with a straight flight path [15]. The laser was operated at 355 nm with a 100 kHz repetition rate, a 50 μm spot size at the specimen, and an average power of 10–12 mW. The specimen was maintained at a base temperature of 160 K. The initial evaporation rate ranged from 100–300 counts per second and was later increased to 300–900 counts per second.

Data evaluation, including mass ranging and tomographic reconstruction, was performed using the software Scito [16]. Initial counts from uncontrolled field evaporation of residual gases, as well as all counts recorded after complete tip rupture, were excluded. The resulting mass spectrum shows a Li signal-to-noise ratio greater than five (see supplementary **Fig. S2**), confirming the reliability of the APT data and the validity of our Li distribution analysis. Depth scaling of the APT reconstruction was optimized using TEM images acquired before APT.

# Results

*Conventional and scanning transmission electron microscopy*

To investigate microstructural modifications induced by lithium implantation while minimizing preparation artifacts, a single TEM sample incorporating both implanted and non-implanted regions was analyzed. The TEM overview image (**Fig. 2a**) provides a cross-sectional view, showing the layered structure of the gold blocking layer, Nb film, and substrate. Thinning windows ~3 µm from the gold edge ensure a well-defined separation between the regions. Virtual dark field imaging of the non-implanted region (**Fig. 2b**), generated from the annular virtual aperture shown in **Fig. 2c**, reveals that the ~280 nm thick pristine Nb film has a polycrystalline columnar microstructure with an average in-plane grain size of 40 ± 5 nm.

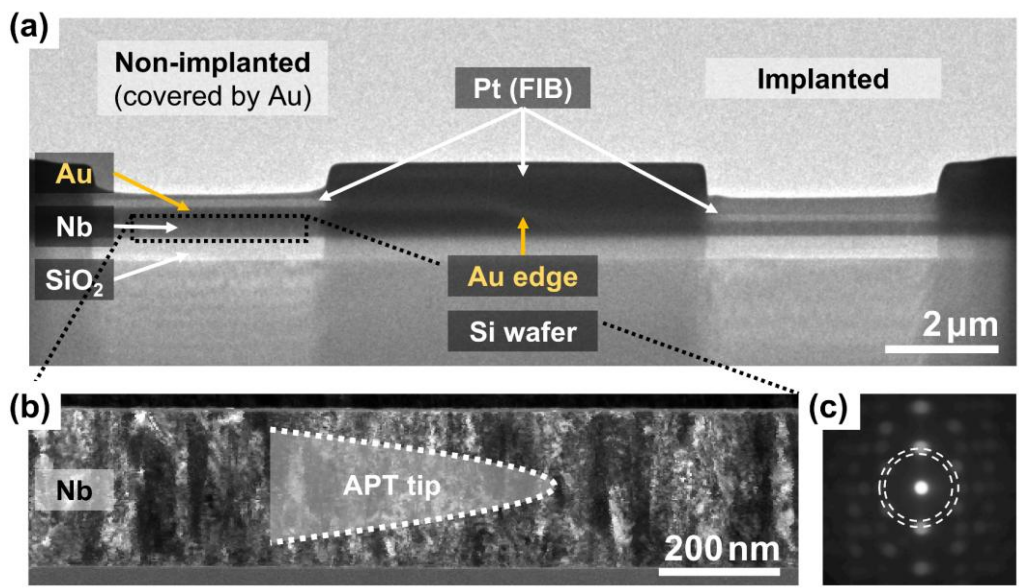

**Figure 2. Cross-sectional TEM and 4D-STEM analysis of the Nb sample.** (**a**) Overview image showing the layered structure of the gold blocking layer, Nb film, and substrate, with thinning windows defining implanted and non-implanted regions. (**b**) Virtual dark field image of the non-implanted region, highlighting the Nb microstructure and orientation of the atom probe tip. (**c**) Averaged diffraction pattern from 4D-STEM data, used to generate (b).

To assess the microstructural effects of Li implantation, 4D-STEM was used to compare the microstructure of Li-free and Li-implanted regions (**Fig. 3**). Different grain orientations were visualized in both regions by virtual STEM imaging, i.e., by RGB overlays in real space (**Fig. 3a, c**) generated by integrating the diffraction intensity within predefined masks (red, green, and blue) including Bragg reflections corresponding to different orientations (**Fig. 3b, d**). These overlays reflect the size, shape and orientation of grains, allowing direct comparison of potential microstructural changes.

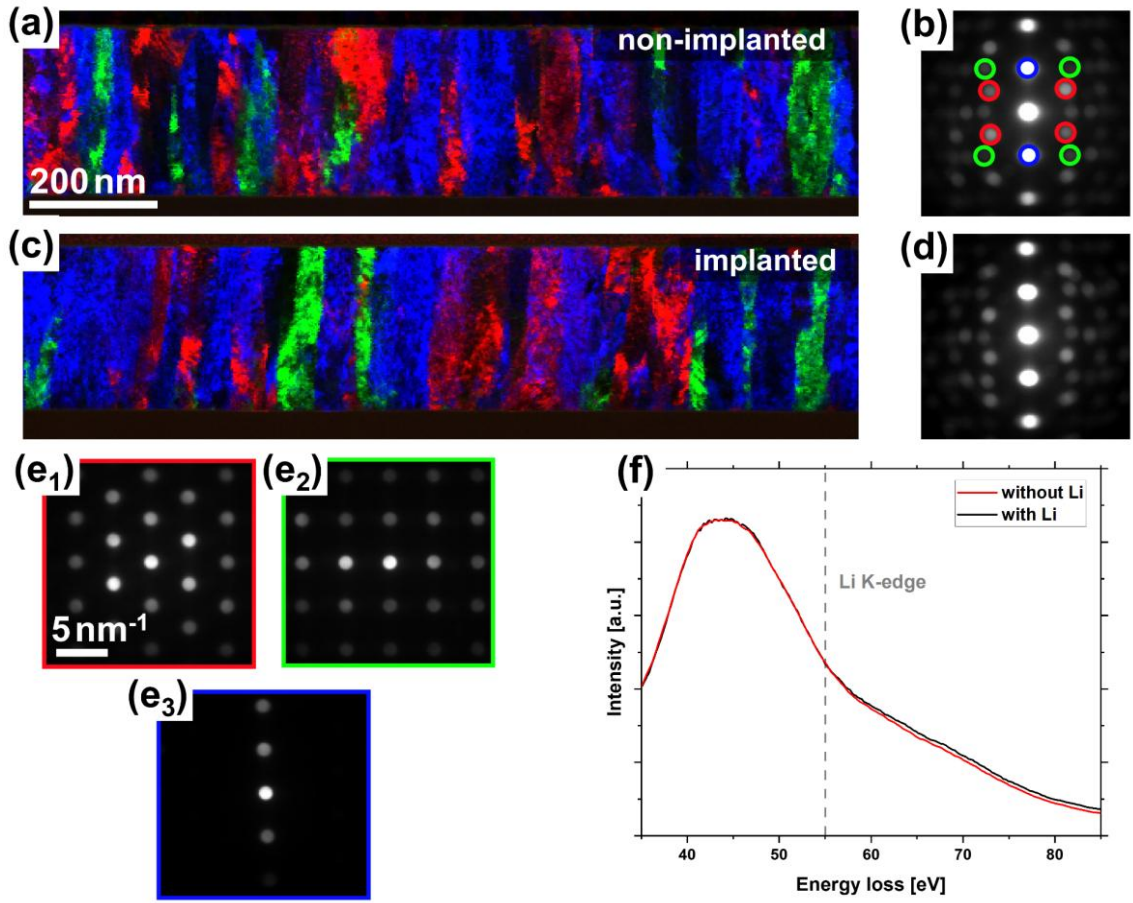

**Figure 3. Grain orientation mapping and diffraction analysis of columnar films without and with implanted Li**. **(a, c)** RGB overlays of differently oriented grains, with (a) showing the Li-free film and (c) the Li-implanted film. The red, green, and blue components correspond to intensity integration within the masks of the respective colors in (b, d) for each individual diffraction pattern from the 4D-STEM data set. **(b, d)** Summed diffraction patterns from all scanning positions in the regions of interests shown in (a) and (c). **(e₁-e₃)** Representative diffraction patterns from different regions of the RGB overlays exhibiting a $[1\bar{1}1]$ zone axis (e$_1$, red), a [001] zone axis (e$_2$, green), and an excitation of the systematic row of the (110) Bragg reflection (e$_3$, blue). f) Electron energy loss spectra summed over the entire film, comparing the Li-free and Li-implanted region.

The successful implantation of Li was confirmed by low-loss EELS spectra (**Fig. 3f**). The spectra were averaged across the regions shown in Fig. 3a, c. and normalized to the maximum of the Nb N-edge at around 45 eV. While no significant differences were observed below the Li K-edge onset at 54.8 eV, the increased intensity at higher energy losses in the Li-implanted region indicates the presence of implanted Li. However, the low signal-to-background ratio and the complex Nb N-edge background shape hindered two-dimensional Li mapping via EELS.

Despite the successful Li implantation, 4D-STEM analysis reveals no detectable microstructural changes. The comparison of averaged diffraction patterns from the Li-free and Li-implanted Nb regions (**Fig. 3b, d**) shows no systematic differences in diffraction spot intensity or position, indicating that the crystallographic structure remains unchanged. Additionally, the absence of a circular diffraction pattern in the implanted Nb region, even in individual diffraction patterns near the surface, suggests no detectable amorphization due to implantation. Grain orientation mapping (**Fig. 3a, c**) further supports this conclusion, showing no significant changes in grain size, shape or orientation between the two regions. Notably, both regions exhibit a pronounced {110} out-of-plane texture, as evidenced by the averaged diffraction pattern (**Fig. 3b**, blue circles) and the corresponding RGB overlay (**Fig. 3a, c**, blue columnar grains).

*Atom probe tomography*

Prior to APT measurement, the sample was characterized by TEM (**Fig. 4a–c**) to enable correlative analysis with APT data. Beneath the amorphous tip apex — likely induced by FIB milling [17,18] — the sample is predominantly crystalline, as confirmed by selected area diffraction (**Fig. 4c**). A distinct grain is visible at ~80 nm depth (**Fig. 4b**).

The tomographic reconstruction (**Fig. 4d**) shows the distribution of Nb (blue) and Li (magenta) atoms. For both elements, an accumulation of atoms is evident that coincides both orientation and position-wise with the observed grain boundary in the TEM images (cf. **Fig. 4b**). The apparent increase in Nb density at the grain boundary is attributed to the local magnification effect [19,20], a known APT artifact at structural defects that occurs even in single-phase materials [21]. On the upside, this effect can also serve as an indicator for grain boundaries, as demonstrated in this case.

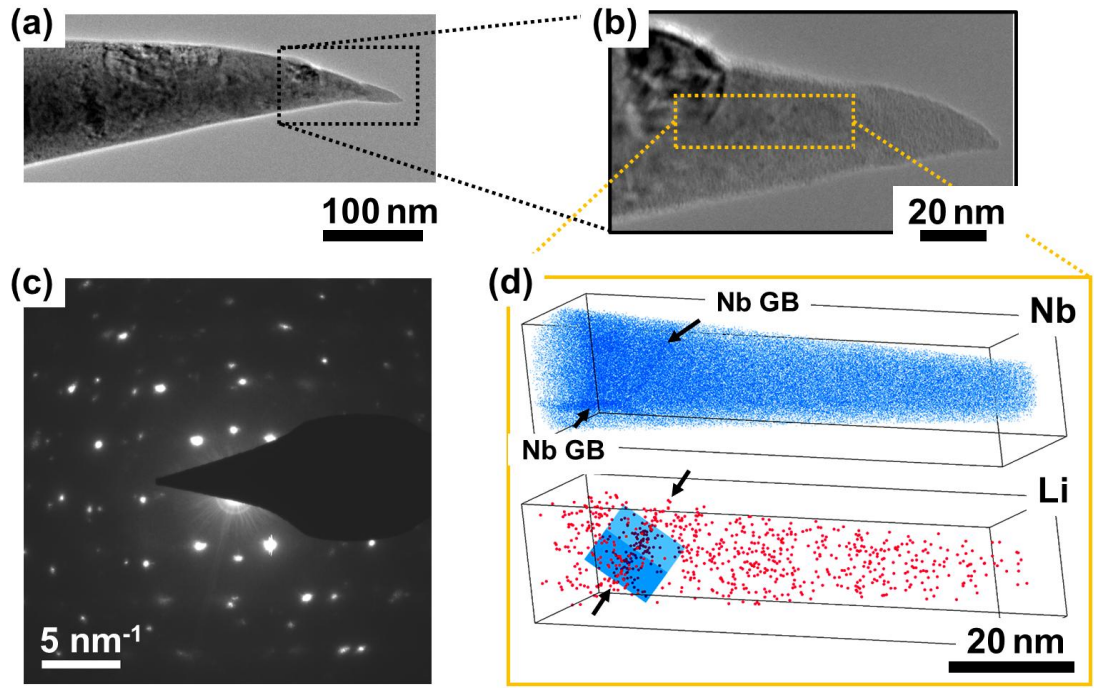

**Figure 4. Correlative TEM and APT analysis of a specimen. (a, b)** Bright-field TEM images of the APT tip, with (b) showing a magnified view of the apex at a 30° tilt relative to (a), where a distinct grain is visible. **(c)** SAEDP from the tip apex, acquired from approximately the same region shown in (b). **(d)** 3D tomographic reconstruction showing the distribution of Li (magenta) and Nb (blue) atoms. The grain boundary, visible in both atom maps, is labeled with black arrows. A virtual cylindrical volume (blue) is positioned perpendicular to the grain boundary for further analysis.

For further analysis, 1D composition profiles were extracted using virtual cylinders (**Fig. 5**). A depth profile from a cylinder centered in the reconstructed volume (4 nm radius, 40 nm length), excluding the grain boundary region, shows an average Li concentration of $(0.60 \pm 0.18)$ at.%, consistent with the intended implantation target of 0.5 at.% Li (**Fig. 5a**). A second cylinder, placed perpendicular to the grain boundary, reveals a pronounced Li segregation peak at $(2.45 \pm 0.41)$ at.%, exceeding the average concentration by more than four standard deviations (**Fig. 5b**).

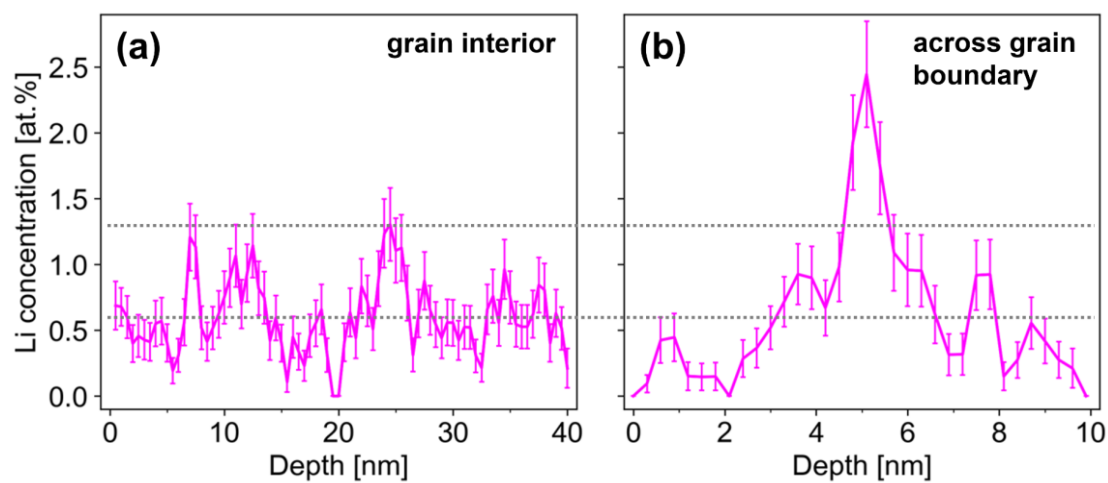

**Figure 5. 1D Li concentration profiles.** **(a)** Depth profile extracted from a cylinder (4 nm radius, 40 nm length, 1 nm slice size, 0.5 nm slice offset) centered in the reconstructed volume, excluding the grain boundary region. **(b)** Grain boundary profile obtained from a cylinder (4.5 nm radius, 10 nm length, 0.6 nm slice size, 0.3 nm slice offset) aligned perpendicular to the grain boundary (shown in blue in Fig. 3), revealing a pronounced grain boundary segregation. Grey dashed lines indicate the average Li concentration of 0.6 at.% and the highest Li concentration measured in the grain interior. Standard deviation error bars were determined following the methodology in [22,23].

The segregation width (chemical width) of the grain boundary, determined from the full width at half maximum (FWHM) of a Lorentzian fit, is ~2.2 nm. However, this width is likely overestimated due to two factors: (1) potential errors from the cylinder's orientation not being perfectly orthogonal to the grain boundary [24] and (2) artifacts inherent to APT, such as the local magnification effect [19,20,25]. As a result, the actual Li concentration at the grain boundary is likely higher than the observed ~2.5 at.% Li.

## Discussion

*Scanning transmission electron microscopy*

The STEM investigations of the effects of lithium implantation revealed no detectable change in microstructure, particularly no amorphization. However, this finding may be limited by the nanosized columnar Nb grains (cf. **Fig. 2b** and **Fig. 3a, c**). Since their grain size is smaller than the lamella thickness (50–100 nm), overlapping in projection hinders the resolution of individual grain boundaries, stacking faults, and dislocations. Furthermore, as STEM cannot detect point defects unless they occur periodically or at high concentrations, no claims about lower-dimensional defects can be made.

*Atom probe tomography*

The APT investigation of Li distribution was constrained by frequent premature sample rupture, with Li segregation at a Nb grain boundary observed in only one measurement. This highlights the need for optimized measurement parameters to improve reproducibility. Increasing the base temperature may enhance specimen yield by reducing rupture likelihood [26]. Notably, even at base temperatures up to 200 K, no artifacts indicative of Li mobility, such as inhomogeneities on the detector screen, were observed [27].

Despite these limitations, the results provide key insights relevant to the BeEST experiment. Atom probe tomography revealed Li segregation at grain boundaries in the Nb matrix, confirmed by correlative TEM (**Fig. 4**). This behavior is attributed to the Gibbs adsorption effect [28,29], where solute segregation at extended defects, such as grain boundaries, reduces defect energy and lowers the system's overall Gibbs free energy. Consequently, Li preferentially accumulates at grain boundaries rather than being uniformly distributed throughout the sample.

*Impact on the BeEST experiment*

The success of the BeEST experiment relies on the performance of STJs. This study suggests that $^7$Li implantation has little impact on the microstructure of Nb-based STJs, up to a dose of at least 0.5 at.% $^7$Li, with similar outcomes expected for $^7$Be implantation. This is significant, as superconducting properties, particularly the critical temperature, can be sensitive to disorder [30]. However, the signal of the BeEST experiment may still be broadened by changes in the chemical environment caused by the implantation process and subsequent electron capture decay.

One example of such a change is the observed Li segregation at Nb grain boundaries (**Fig. 5b**), which affects the local electronic structure of lithium atoms in the Nb layer. Density functional theory (DFT) simulations of Li in Ta indicate that such segregation can shift the Li 1s level by up to 2 eV [8]. Given the atomic and chemical similarities between Ta and Nb—including their crystal structure, electronic structure, and density of states—similar energy shifts and segregation tendencies are expected for Li in Nb [31]. While Li segregation at grain boundaries may contribute to the observed peak broadening, it does not preclude the use of Nb-based STJ detectors in the BeEST sterile neutrino search.

## Conclusion

This study evaluates the impact of $^7$Li implantation on Nb films as a potential sensor material for the BeEST sterile neutrino search [6]. STEM analysis reveals no detectable microstructural changes, particularly no amorphization, suggesting that implantation of up to ~0.5 at.% $^7$Li is unlikely to degrade Nb-STJ performance. APT confirms some segregation of $^7$Li at Nb grain boundaries, attributed to the Gibbs adsorption effect. While this alters the local chemical environment, density functional theory (DFT) simulations suggest that the resulting energy shift is not expected to significantly broaden the recoil spectrum. Overall, these findings indicate that Nb remains a viable sensor material for future phases of the BeEST experiment.

## Acknowledgements

We thank Prof. Guido Schmitz for granting access to the atom probe and other laboratory equipment at the Institute for Materials Science in Stuttgart. We also gratefully acknowledge the use of equipment at the "Collaborative Laboratory and User Facility for Electron Microscopy" (CLUE – www.clue.physik.uni-goettingen.de). Part of this work was performed under the auspices of the U.S. DOE by LLNL under Contract No. DEAC52-07NA27344.

Funding from the Deutsche Forschungsgemeinschaft (DFG) within the Research Group „Energy Landscapes and Structure in Ion Conducting Solids" under Project Number 428906592 is gratefully acknowledged.

## Disclosure of interest

The authors report there are no competing interests to declare.

# Supplementary Materials

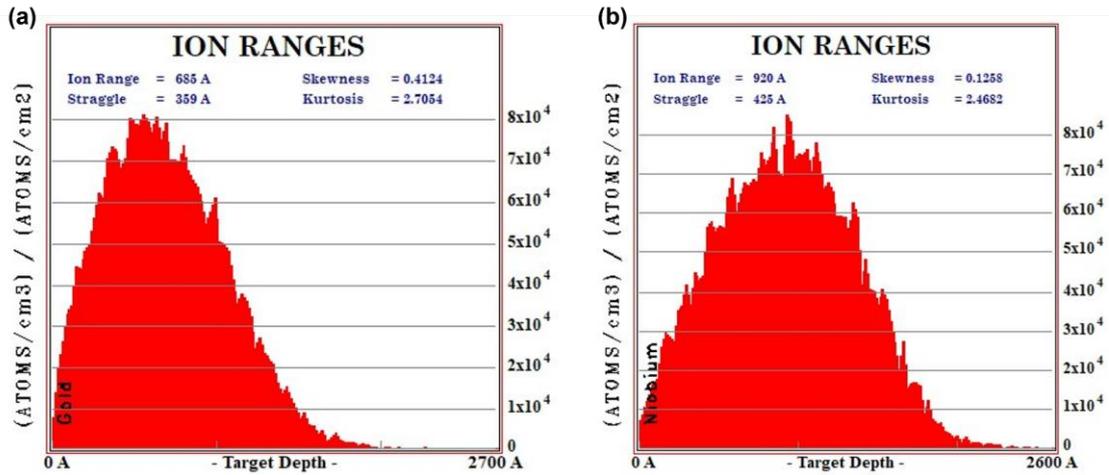

**Figure S1. SRIM simulations of $^7$Li** with a kinetic energy of 30 keV in (a) 270 nm of Au and (b) 260 nm of Nb. The SRIM-2013 software package was used [10].

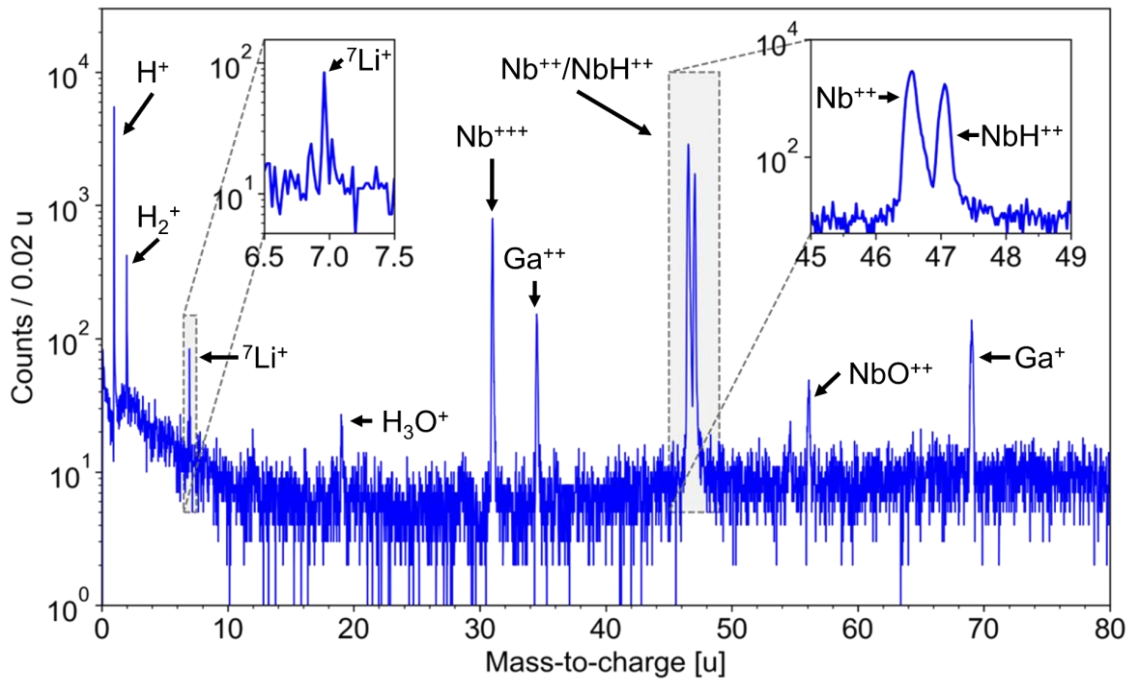

**Figure S2. APT mass spectrum** from an inner cylinder (4 nm radius, 40 nm length), with insets highlighting the signal-to-noise ratio greater than five for the $^7$Li$^+$ peak and the significant evaporation of Nb as NbH$^{++}$ ions. The detected hydrogen and H$_3$O$^+$ signals are attributed to residual gas in the vacuum chamber, while Ga originates from the FIB milling process.

16